\documentclass[fleqn,10pt]{wlscirep}
\usepackage[utf8]{inputenc}
\usepackage[T1]{fontenc}
\usepackage{bm}
\usepackage{float}
\usepackage{algorithm} 
\usepackage{algpseudocode} 
\usepackage{mathtools}
\usepackage{algcompatible}
\usepackage{accents}

\usepackage{float}
\usepackage{pifont}
\usepackage{multirow}
\usepackage{listings}
\usepackage{listings}

\title{MDDC: An R and Python Package for Adverse Event Identification in Pharmacovigilance Data}

\author[1,+]{Anran Liu}
\author[1,+]{Raktim Mukhopadhyay}
\author[1,*]{Marianthi Markatou}

\affil[1]{University at Buffalo, Department of Biostatistics, Buffalo, 14214, USA}

\affil[*]{Correspondence and requests should be addressed to M.M. (email: markatou@buffalo.edu)}

\affil[+]{Authors contributed equally to this work}

\begin{abstract}
The safety of medical products continues to be a significant health concern worldwide. Spontaneous reporting systems (SRS) and pharmacovigilance databases are essential tools for postmarketing surveillance of medical products. Various SRS are employed globally, such as the Food and Drug Administration Adverse Event Reporting System (FAERS), EudraVigilance, and VigiBase. In the pharmacovigilance literature, numerous methods have been proposed to assess product - adverse event pairs for potential signals. In this paper, we introduce an \texttt{R} and \texttt{Python} package that implements a novel pattern discovery method for postmarketing adverse event identification, named Modified Detecting Deviating Cells (MDDC). The package also includes a data generation function that considers adverse events as groups, as well as additional utility functions. We illustrate the usage of the package through the analysis of real datasets derived from the FAERS database.
\end{abstract}

\keywords{adverse events, Modified Deviating Data Cells (MDDC) algorithm, pattern discovery, pharmacovigilance, public health, software}

\makeatletter
\newenvironment{breakablealgorithm}
  {
   \begin{center}
     \refstepcounter{algorithm}
     \hrule height.8pt depth0pt \kern2pt
     \renewcommand{\caption}[2][\relax]{
       {\raggedright\textbf{\ALG@name~\thealgorithm} ##2\par}%
       \ifx\relax##1\relax 
         \addcontentsline{loa}{algorithm}{\protect\numberline{\thealgorithm}##2}%
       \else 
         \addcontentsline{loa}{algorithm}{\protect\numberline{\thealgorithm}##1}%
       \fi
       \kern2pt\hrule\kern2pt
     }
  }{
     \kern2pt\hrule\relax
   \end{center}
  }
\makeatother

\begin{document}

\flushbottom
\maketitle
%
%
\thispagestyle{empty}



\section{Introduction}

Adverse events from medical products, such as drugs, therapeutic biologics, or medical devices are a serious concern in healthcare because they can result in hospitalizations and deaths. Clinical trials are the main sources of safety information pre-licensure. However, clinical trials have limitations such as relatively small sample sizes and short duration. Rare but serious adverse events are likely not to be noticed until the medical product is on the market. Therefore, continual monitoring of medical products after their introduction in the market is crucial for ongoing safety monitoring. 

Spontaneous reporting systems (SRS) and pharmacovigilance databases are essential tools for postmarketing surveillance of medical products. Various SRS are used worldwide, such as Food and Drug Administration (FDA) Adverse Event Reporting System (FAERS), the European Medicines Agency's EudraVigilance, and World Health Organization's VigiBase. Additionally various methodologies exist for analyzing data from SRS in pharmacovigilance, and several packages have been developed for the analysis of pharmacovigilance data. These methods vary from traditional disproportionality analysis, including Proportional Reporting Ratios (PRR), Reporting Odds Ratios (ROR), to Bayesian methods such as Multi-item Gamma Poisson Shrinker (MGPS) and Bayesian Confidence Propagation Neural Network (BCPNN). The \texttt{R} package \texttt{PhViD} \cite{ahmed2016phvid} implements several key methodologies that are widely used, including PRR, ROR, Gamma Poisson Shrinker (GPS) and BCPNN. The package \texttt{vigipy} \cite{Beery2024} also implements and extends the methods found in \texttt{PhVid} in the \texttt{Python} ecosystem. The \texttt{R} package \texttt{pvm}\cite{Dijkstra2020} implements several methods for AE identification, such as ROR, PRR, BCPNN, and Penalized Regression, to name a few. The \texttt{openEBGM}\cite{canida2017openebgm} package implements the MGPS method. The likelihood ratio test (LRT) based method introduced by Huang et al. \cite{Huang2011}, and its subsequent refinements, extensions and reparametrizations, offer a comprehensive suite of rigorous statistical techniques for addressing pharmacovigilance data analysis.\cite{Huang2011, Huang2017, Zhao2018, Ding2020, Chakraborty2022} These methods, which assume either the Poisson or Zero-Inflated Poisson models, illustrate the association between adverse events and medical products through the use of likelihood ratio test statistics. The \texttt{R} package \texttt{pvLRT}\cite{pvlrt2023} implements a reparametrization of the LRT method that unifies the Poisson and ZIP models with extensions of testing multiple drugs simultaneously. Furthermore it incorporates the original LRT method by Huang et al.\cite{Huang2011}. Other packages encoding methods for medical product safety include the \texttt{R} packages \texttt{AEenrich}\cite{Li2022}, \texttt{sglr}\cite{Narasimhan2012}, and \texttt{Sequential}\cite{Silva2021} for sequential testing based vaccine safety, and \texttt{mds}\cite{Chung2020} for data preprocessing in medical devices surveillance. 

Regarding pattern discovery methods in adverse events identification, Kulldorf et al.\cite{kulldorff2003tree, kulldorff2013drug} proposed a data mining method \texttt{TreeScan} for disease surveillance and post-marketing drug safety surveillance using a tree-based scan statistic in hierarchically structured data. The problem of pattern discovery with statistical performance guarantees in the context of pharmacovigilance has not been sufficiently discussed in the literature. We present software developed in \texttt{R} and \texttt{Python} that encodes a novel statistical pattern discovery method proposed by Liu et al.\cite{Liu2024}. This method, called Modified Detecting Deviating Cells (MDDC) algorithm is able to identify adverse events associated with count data on pairs of (adverse event, medical product) that is organized as an $I\times J$ contingency table. 

In Table \ref{tab:comparision}, we present a brief comparison of the various packages. The comparison is not intended in terms of performance, but with respect to implementation aspects such as programming language, active development, package repository, unit tests and coverage and continuous integration. 

The article is subsequently organized as follows. Section \ref{sec:methodlogy} provides the required background on the proposed methodology. Section \ref{sec:package-desc} discusses in detail the structure of the packages in both \texttt{R} and \texttt{Python}. Section \ref{sec:example} exemplifies the application of our methods on a beta blocker dataset derived from FAERS, while Section \ref{sec:performance} presents simulations with the aim of studying the performance of our package, and suggest extensions of our data generation algorithm. Section \ref{sec:discussion} offers discussion and conclusions.


\begin{table}[H]
\centering
\resizebox{\textwidth}{!}{%
\begin{tabular}{|
>{\columncolor[HTML]{C0C0C0}}c |c|c|c|c|c|}
\hline
\textbf{Packages/Software} & \cellcolor[HTML]{C0C0C0}\textbf{\begin{tabular}[c]{@{}c@{}}Programming \\ Language\end{tabular}} & \cellcolor[HTML]{C0C0C0}\textbf{\begin{tabular}[c]{@{}c@{}}Active \\ Development \\ (2023)\end{tabular}} & \cellcolor[HTML]{C0C0C0}\textbf{\begin{tabular}[c]{@{}c@{}}Package/Software\\ Repository\end{tabular}} & \cellcolor[HTML]{C0C0C0}\textbf{\begin{tabular}[c]{@{}c@{}}Unit Tests \\ (Coverage \%)\end{tabular}} & \cellcolor[HTML]{C0C0C0}\textbf{\begin{tabular}[c]{@{}c@{}}Continuous \\ Integration\end{tabular}} \\ \hline
\texttt{\textbf{PhVid}} & \texttt{R} & \textcolor{red}{\xmark} & CRAN* & \textcolor{red}{\xmark} & \textcolor{red}{\xmark} \\ \hline
\texttt{\textbf{openEBGM}} & \texttt{R} & \textcolor{green}{\cmark} & CRAN & \textcolor{green}{\cmark}(68\%) & \textcolor{red}{\xmark} \\ \hline
\texttt{\textbf{pvLRT}} & \texttt{R} & \textcolor{green}{\cmark} & CRAN & \textcolor{red}{\xmark} & \textcolor{red}{\xmark} \\ \hline
\texttt{\textbf{AEenrich}} & \texttt{R} & \textcolor{red}{\xmark} & CRAN & \textcolor{red}{\xmark} & \textcolor{red}{\xmark} \\ \hline
\texttt{\textbf{sglr}} & \texttt{R} & \textcolor{red}{\xmark} & CRAN & \textcolor{red}{\xmark} &  \textcolor{red}{\xmark}\\ \hline
\texttt{\textbf{Sequential}} & \texttt{R} & \textcolor{red}{\xmark} & CRAN & \textcolor{red}{\xmark}  &  \textcolor{red}{\xmark}\\ \hline
\texttt{\textbf{vigipy}} & \texttt{Python} & \textcolor{red}{\xmark} & GitHub & \textcolor{green}{\cmark}  &  \textcolor{red}{\xmark}\\ \hline
\texttt{\textbf{pvm}} & \texttt{R} & \textcolor{red}{\xmark} & GitHub & \textcolor{red}{\xmark}  &  \textcolor{red}{\xmark}\\ \hline
\texttt{\textbf{mds}} & \texttt{R} & \textcolor{red}{\xmark} & CRAN & \textcolor{green}{\cmark} (87.5 \%)  &  \textcolor{red}{\xmark}\\ \hline
\cellcolor[HTML]{C0C0C0} & \texttt{C++}/\texttt{Java} &  \textcolor{green}{\cmark} & CRAN & \textcolor{green}{\cmark} & -- \\ \cline{2-6} 
\multirow{-2}{*}{\cellcolor[HTML]{C0C0C0}\texttt{\textbf{TreeScan}}} & \texttt{R} & \textcolor{green}{\cmark} & GitLab & \textcolor{red}{\xmark} & \textcolor{red}{\xmark} \\ \hline 

\cellcolor[HTML]{C0C0C0} & \texttt{R}/\texttt{C++} &  \textcolor{green}{\cmark} & CRAN & \textcolor{green}{\cmark} (99.1 \%) & \textcolor{green}{\cmark} \\ \cline{2-6} 
\multirow{-2}{*}{\cellcolor[HTML]{C0C0C0}\texttt{\textbf{MDDC}}} & \texttt{Python}/\texttt{C++} & \textcolor{green}{\cmark} & PyPI & \textcolor{green}{\cmark} (98.5 \%) & \textcolor{green}{\cmark} \\ \hline 
\end{tabular}%
}

\caption{Comparison of various relevant packages.\\
\textbf{Active Development (2023)} is determined by the release of any new version of the package/software between January 1, 2023, and December 31, 2023.\\
*: The package has been archived from CRAN at the time of writing this paper.
}
\label{tab:comparision}
\end{table}




\section{Methodology}
\label{sec:methodlogy}

We provide a brief review of the proposed MDDC algorithm for adverse event identification. For a detailed description and discussion of this algorithm see Liu et al \cite{Liu2024}.

Individual Case Safety Reports (ICSRs) are important tools used in monitoring the safety of medical products; the FAERS reporting system, the database where these reports are submitted, receives millions of reports each year. If interest centers on a specific time period, then ICSRs can be extracted from the safety database and depicted as an $I \times J$ contingency table, with cell counts $n_{ij}, i = 1, 2,...,I, j=1,2,...,J$ indicating the total number of reported cases corresponding to the $i$th AE and $j$th drug. The total number of reported AEs equals to $I$, while $J$ denotes the total number of products, for example drugs. We are interested in identifying which AE - drug pairs are signals. The signals refer to potential adverse events that may be caused by a drug. In the contingency table setting, the signals refer to the cells with $n_{ij}$ abnormally higher than their expected values. 

Liu et al.\cite{Liu2024}, inspired by Rousseeuw and Bossche\cite{Rousseeuw2018}, significantly modified and expanded the work of these authors to better suit the discrete nature of pharmacovigilance's measurement scale. Our Modified Detecting Deviating Cells (MDDC) algorithm has the following characteristics: 1) it is easy to compute; 2) it considers AE relationships; 3) it depends on data driven cutoffs; 4) it is independent of the use of ontologies. The MDDC algorithm involves five steps, with the first two steps identifying univariate outliers via cutoffs, the next three steps evaluating the signals via the use of AE correlations. The five steps of the MDDC method are described in Algorithm~\ref{mddc_alg}. 

\newpage

\begin{breakablealgorithm}
\caption{Modified Detecting Deviating Cells (MDDC) algorithm 
\label{mddc_alg}}
\begin{algorithmic}[1]
        \item For each cell in the $I \times J$ contingency table compute the standardized Pearson residual:
        $$e_{ij} = \frac{n_{ij}-(n_{i\bullet}n_{\bullet j}/n_{\bullet \bullet})}{\sqrt{(n_{i\bullet}n_{\bullet j}/n_{\bullet \bullet})(1-n_{i\bullet}/n_{\bullet \bullet})(1-n_{\bullet j}/ n_{\bullet\bullet})}}.$$

        \item Separate all the cells in the table into two groups, the non-zero cells of which $n_{ij}> 0$, and the observed zero cells of which $n_{ij}=0$. Let $\{e^+_{ij}\}$ denote the $\{e_{ij}\}$ of all the non-zero cells, and let $\{e^0_{ij}\}$ denote the $\{e_{ij}\}$ of all the observed zero cells. Define a matrix $\mathbf{U}$ with entries

$$
u_{ij} = 
\begin{cases}
e^+_{ij}, \ \text{ if } |e^+_{ij}|\leq c^+_{univ,j}\\
\text{NA}, \text{ if } |e^+_{ij}|> c^+_{univ,j}\\
e^0_{ij}, \ \text{ if } e^0_{ij}\geq c^0_{univ,j}\\
\text{NA}, \text{ if } e^0_{ij}< c^0_{univ,j}
\end{cases}.
$$We need to identify the cutoff points $c^+_{univ,j}$, $c^0_{univ,j}$. To obtain $c^+_{univ,j}$, we propose two methods; Tukey's boxplot and a Monte Carlo (MC) method. For the $c^0_{univ,j}$ cutoff, that is, the cutoff associated with the zero cells, we propose the use of Tukey's boxplot statistic defined as 
$c_{univ,j}^0=Q1(\{e^0_{ij}\})-1.5\times IQR(\{e^0_{ij}\}),$ where $Q1$ and $Q3$ are the first and third quartile of the standardized Pearson residuals associated with the zero cells, $IQR$ is the interquartile range. The upper limit of the Tukey boxplot statistic defined as $c_{univ,j}^+=Q3(\{e^+_{ij}\})+1.5\times IQR(\{e^+_{ij}\})$is one of the choices to obtain $c_{univ,j}^+$. Algorithm~\ref{alg_boxplot_c} provides for adaptive selection of $c$. We provide an alternative way using Monte Carlo simulation to identify $c_{univ,j}^+$ in Algorithm~\ref{bootstrap_alg}. The cells with $e^+_{ij} > c^+_{univ,j}$ are labeled as ``upper outliers''.

\vspace{0.5em}
    
        \item For any two AE rows $i\neq k$, compute their Pearson correlation, $cor_{ik} = Corr(\tilde{u}_{i}, \tilde{u}_{k})$, where $\tilde{u}_{i} = (u_{i1}, u_{i2}, ..., u_{iJ})$ and $\tilde{u}_{k} = (u_{k1}, u_{k2}, ..., u_{kJ})$. Let $c_{corr}$ be the correlation threshold. For the $i$-th AE, if $|cor_{ik}| \geq c_{corr}$ for $k\neq i$, then AE $k$ is called a ``connected'' AE to AE $i$.  
       \vspace{0.5em}
        
        \item Obtain predicted values for each cell based on the connected AEs. Suppose there are $m$ connected AEs for AE $i$, that is, $k = k_1,k_2,...,k_m$, with correlations $cor_{ik_1}, cor_{ik_2},...,cor_{ik_m}$. For each connected AE $k$, fit a simple linear regression with intercept $\alpha_{ik}$ and slope $\beta_{ik}$ using $\tilde{u}_{i}$ as the response variable and $\tilde{u}_{k}$ as the explanatory variable. The fitted value for AE $i$ based on AE $k$ $\{\hat{u}_{ikj}\}$ are then obtained as $\{\alpha_{ik} + \beta_{ik}u_{kj}\}$. The fitted value for AE $i$ based on all the connected AEs are obtained as a weighted mean with weights $w_{ik} = \frac{|cor_{ik}|}{\sum_{l=k_1}^{k_m}|cor_{il}|}$, and the weighted mean is $\hat{u}_{ij} = w_{ik_1}\hat{u}_{ik_1j}+w_{ik_2}\hat{u}_{ik_2j}+...+w_{ik_m}\hat{u}_{ik_mj}.$ 
      \vspace{0.5em}
  
        \item Compute $r_{ij} = \left[(e_{ij}-\hat{u}_{ij})-A_j\right]/{\sqrt{B_j}},$ the computation is over all $i$ within drug $j$ for which neither $e_{ij}$ or $\hat{u}_{ij}$ is NA, where $A_j = \frac{1}{I_j}\sum_{i\in I_j}(e_{ij}-\hat{u}_{ij})$, $B_j= \frac{1}{I_j}\sum_{i\in I_j}\left[(e_{ij}-\hat{u}_{ij})-A_j\right]^2$, $I_j$ denotes the row indices of cells in drug$_j$ that neither $e_{ij}$ or $\hat{u}_{ij}$ is NA. Calculate the upper tail probability of $r_{ij}$ in the standard normal distribution, which is used as the p-value for each cell. Obtain adjusted p-values via Benjamini-Hochberg procedure to control false discovery rate. The second set of signals are obtained as the cells with adjusted p-values less than 0.05.
    \vspace{0.5em}
        
\end{algorithmic}
\end{breakablealgorithm}

In step 2, the cutoff point $c^+_{univ,j}$ can be identified using two methods. One method is via the upper limit of Tukey's Boxplot statistics, which is defined as $Q3+c\times IQR$. In the upper limit, $Q3$ is the third quartile of the standardized Pearson residuals associated with the non-zero cells and $IQR$ stands for interquartile range. Tukey suggested using the coefficient $c=1.5$.\cite{tukey1977exploratory} However, this value is most
appropriate when the data follows a normal distribution. The boxplot coefficient $c$ can be adjusted to
accommodate different data distributions. Algorithm~\ref{alg_boxplot_c} offers an adaptive approach to determine the optimal boxplot coefficient $c$ that facilitates accurate identification of adverse events based on the concept of False Discovery Rate (FDR), defined in Algorithm \ref{alg_boxplot_c}.

\begin{algorithm}[H]
\caption{Adaptive approach to determine the optimal boxplot coefficient $c$}\label{alg_boxplot_c}
    \begin{enumerate}
\item For a given contingency table of dimension $I \times J$ calculate the $n_{\cdot \cdot}$, and the $\undertilde{p} = p_{11}, p_{12}, \ldots p_{IJ}$.

\item Generate a large number of $I\times J$ tables $r=1,...,R$ under the assumption of independence from multinomial distribution using the $n_{\cdot \cdot}$ and $\undertilde{p}$ determined in Step 1.
\vspace{0.5em}
\item Compute the standardized Pearson residuals of each table.
\vspace{0.5em}
\item Compute the upper limits with $c=1.5$, and compute false discovery rate (FDR) as $\frac{1}{R}\sum_{r=1}^R (FP^r)/(FP^r+TP^r)$, where $FP^r$ denotes the number of false positives in the $r$th table, $TP^r$ denotes the number of true positives. Under independence of rows and columns, $TP^r=0$.
\vspace{0.5em}
\item If FDR$\leq 0.05$, stop; if FDR $> 0.05$, add a small value to $c$ such as $c=c+0.1$ and repeat step 4, until we find the best $c$ for which FDR is less than or equal to 0.05.
    \end{enumerate}
    \end{algorithm}

The next algorithm, Algorithm~\ref{bootstrap_alg}, provides an alternative way to identify cutoffs for the second step of Algorithm~\ref{mddc_alg}. It is based on the idea of computing the standardized Pearson residuals under the assumption of independence of rows and columns. In our pharmacovigilance context, independence between rows and columns indicates that signals of adverse events are not associated with the medical products under consideration. We describe the various steps in Algorithm~\ref{bootstrap_alg}. The standardized Pearson residual measures the difference between the observed cell count and the expected cell count in a contingency table, which is often time used in the context of chi-squared tests of independence. However, in the situations where the expected cell counts are small (less than 5), the residuals may become unstable and unreliable\cite{Agresti2012}. To avoid misleading results, in step 4 of Algorithm~\ref{bootstrap_alg}, the maximum of the standardized Pearson residuals is taken on the cells with a count of at least 6 by multiplying the residuals with the indicator function $\bm{1}\{n_{ij}>5\}$.

\begin{algorithm}[H]
\caption{Monte Carlo simulation method for obtaining the cutoff value $c_{univ,j}^+$ in Step 2 of the MDDC method }\label{bootstrap_alg}
\begin{algorithmic}[1]
    
         \item Obtain the marginals $n_{1\bullet},...,n_{I\bullet}, n_{\bullet 1},...,n_{\bullet J}$ from the original $I\times J$ contingency table. 
         \vspace{0.5em}

         \item Under the assumption of no association between drugs and AEs (i.e., independence of rows and columns), compute cell probabilities$$p_{ij} = (\frac{n_{i\bullet}}{n_{\bullet \bullet}})(\frac{n_{\bullet j}}{n_{\bullet \bullet}}),$$where $i = 1,...,I$ and $j = 1,...,J$.
         \vspace{0.5em}

         \item Generate 10,000 $I\times J$ contingency tables with the above specified marginals and cell probabilities $\{p_{ij}\}$ through multinomial distribution$$(n_{11}, n_{12},...,n_{IJ}) \sim \text{Multinomial}(n_{\bullet \bullet}, \bm{p}),$$ where $\bm{p} = (p_{11}, p_{12},...,p_{IJ})^T$.
         \vspace{0.5em}
         
         \item For the $r$-th simulated table, $r=1,...,10000$, compute $e_{ij}$ for all the cells in the table, and obtain
         $$m_{j,r} = \max_{1\leq i \leq I} e_{ij}\times \bm{1}\{n_{ij}>5\}$$ for $j = 1,...,J$. For each drug $j$, this will provide $m_{j,1}, m_{j,2},...,m_{j,10000}$.
         \vspace{0.5em}

         \item For each drug $j$, obtain the cutoff value $c_{univ,j}^+$ as the 95-th quantile by ordering $m_{j,1}, m_{j,2},...,m_{j,10000}$ from smallest to largest.   
    \end{algorithmic}
\end{algorithm}

After obtaining the cutoff value $c_{univ,j}^+$ via Algorithm~\ref{bootstrap_alg}, we can determine which cells are potential AE signals by comparing the observed residuals $\{e_{ij}^+\}$ with $c_{univ,j}^+$ as described in the step 2 of the MDDC algorithm. For cells with a count less than 6, we can still get a preliminary judgment by comparing their residuals with $c_{univ,j}^+$.  To address the case where the number of reports is less than 6, we perform Fisher's exact tests on the associated $2\times 2$ tables. To ensure a better control of the false discovery rate, instead of comparing the $\{e_{ij}^+\}$ with $c_{univ,j}^+$, one can use the adjusted p-values obtained via the Benjamini-Hochberg procedure. Cells with adjusted p-values less than 0.05 will then be identified as AE signals. 

In addition, our software includes a data generation function specifically designed for simulating pharmacovigilance datasets. In the current literature, there is no data generation method addressing the embedding of correlations between standardized Pearson residuals. This function utilizes standardized Pearson residuals and allows for the consideration of AEs as clusters, embedding AE correlations into the simulation. The detailed steps of this data generation process are presented in Algorithm~\ref{data_generate_alg_new}.

\begin{breakablealgorithm}
\caption{A data generation algorithm considering correlated AEs}\label{data_generate_alg_new}
Suppose there are $G$ groups (clusters) of AEs in the contingency table, the number of AEs in group $g$ is $c_g$, $g = 1,...,G$.
\begin{algorithmic}[1]

\item For each group $g$, generate the ${e_{ij}}$ from a multivariate normal distribution. Suppose the row indices of the AEs in group $g$ are $1^*, 2^*,...,c_g^*$, generate
$$\begin{pmatrix}
e_{1^*, j}\\
e_{2^*, j}\\
\vdotswithin{}\\
e_{c_g^*,j}\\

\end{pmatrix}\sim N\left(\bm{\mu} = \begin{pmatrix}
0\\
0\\
\vdotswithin{}\\
0
\end{pmatrix}, \Sigma = \begin{pmatrix}
    1 & \rho_{1^*,2^*} & \ldots &\rho_{1^*,c_g^*}\\
    \rho_{2^*,1^*} & 1 & \ldots &\rho_{2^*,c_g^*}\\
    \vdotswithin{} & \vdotswithin{}&  & \vdotswithin{}\\
    \rho_{c_g^*,1^*} & \rho_{c_g^*,2^*} & \ldots&1
\end{pmatrix}\right),
$$for $j = 1,...,J$, and $g = 1,...,G$. $\rho_{i^*,l^*}$ denotes the correlation between AEs $i^*$ and $l^*$ within group $g$, and $\rho_{i^*,l^*} \in [0,1], \quad \forall i^*, l^* =1,..., c_g^*$.
\vspace{0.5em} 

In this step, we obtain $\{e_{ij}\}$ for all the cells in the $I\times J$ table by embedding the between AE correlation $\rho_{i^*,l^*}$ for group $g$.
\vspace{0.5em} 

\item For each cell compute $x_{ij} = e_{ij}\sqrt{E_{ij}\lambda_{ij}(1-p_{i \bullet})(1-p_{\bullet j})}+E_{ij}\lambda_{ij},$ where $\lambda_{ij}$ denotes the signal strength; $\lambda_{ij} > 1$ indicates the cell is a signal and $\lambda_{ij} = 1$ indicates the cell is non-signaled. The cell count $n_{ij}$ is obtained as 
$$
n_{ij} = \begin{cases}
    \left\lfloor x_{ij}\right\rceil, \text{  if  } x_{ij} \geq 0,\\
     \\
    \ \ \ 0, \text{  if  } x_{ij}<0.
\end{cases}
$$where $\left\lfloor x \right\rceil$ is the nearest integer function that takes as input a real number $x$ and gives as output the nearest integer to the input real number. If the fractional part of the number is exactly 0.5, the function rounds to the nearest even integer (round half to even). The value assigned to $n_{ij}$ should be round to an integer as it is a count. If a negative value is obtained, it should be replaced by 0. 

\vspace{0.5em} 

In this step, we obtain $\{n_{ij}\}$ of all the cells in the contingency table.

\end{algorithmic}
\end{breakablealgorithm}

\section{Package Description}
\label{sec:package-desc}

\subsection*{Package Structure}

\begin{figure}[H]
    \centering
    \includegraphics[width=0.8\linewidth]{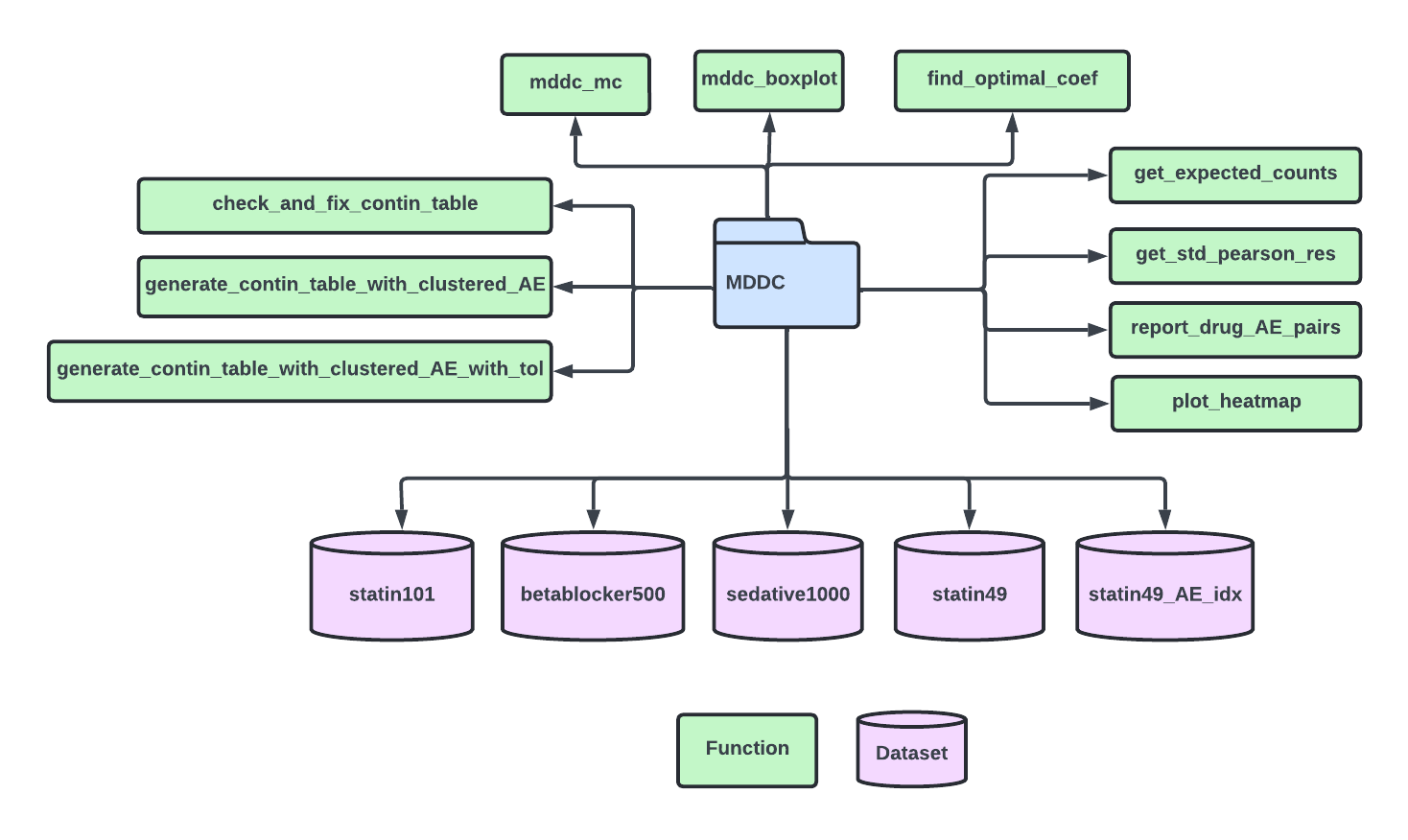}
    \caption{The various functions and datasets available in the \texttt{MDDC} package in \texttt{R}. The folder (blue) in the center depicts the \texttt{MDDC} package. The rectangles (green) represent the various functions which are available in the package. Additionally, the package contains datasets shown with the data icon (lavender-colored).}
    \label{fig:R-MDDC}
\end{figure}

The \texttt{MDDC} package in \texttt{R} and \texttt{Python} implements the MDDC algorithm described in Algorithm~\ref{mddc_alg} alongwith additional functions which are aimed at streamlining the process of identiying AEs. The organization of the \texttt{R} package is shown in Figure \ref{fig:R-MDDC}. 

The main functionality of the package is available in the functions \texttt{mddc\_mc} and \texttt{mddc\_boxplot}. These functions implement the Monte-Carlo and Boxplot methods used for cutoff identification of the MDDC algorithm respectively. The pre-processing function \texttt{check\_and\_fix\_contin\_table} ensures that the input contingency table to \texttt{mddc\_mc} and \texttt{mddc\_boxplot} is correctly formatted. The function \texttt{find\_optimal\_coef} performs a grid search to determine the optimal adaptive boxplot coefficient for each column of the input contingency table, ensuring that a specified target false discovery rate is met. The determined coefficients can also be input to the \texttt{mddc\_boxplot} function. 

The post-processing function \texttt{report\_drug\_AE\_pairs} tabulates the identified signals. Additionally, the function \texttt{generate\_contin\_table\_with\_clustered\_AE} implements the data generation algorithm described in Algorithm \ref{data_generate_alg_new}. A modification of the data generation algorithm (Algorithm \ref{data_generation_with_tol}) and its implementation is presented in the function\\ \texttt{generate\_contin\_table\_with\_clustered\_AE\_with\_tol} to ensure that the total number of reports in the contingency table remain the same in each simulation run. The package also includes the \texttt{plot\_heatmap} function for visualizing identified signals as a heatmap, along with utility functions like \texttt{get\_expected\_counts} to extract expected counts and \texttt{get\_std\_pearson\_res} to calculate standardized Pearson residuals. The package incorporates five datasets (\texttt{statin49}, \texttt{statin49\_AE\_idx}, \texttt{statin101}, \texttt{betablocker500}, and \texttt{sedative1000}) derived from FAERS. The \texttt{R} package can be found on GitHub at \url{https://github.com/niuniular/MDDC/},  with its documentation accessible at \url{https://niuniular.github.io/MDDC/}. 

\begin{figure}
    \centering
    \includegraphics[width=\linewidth]{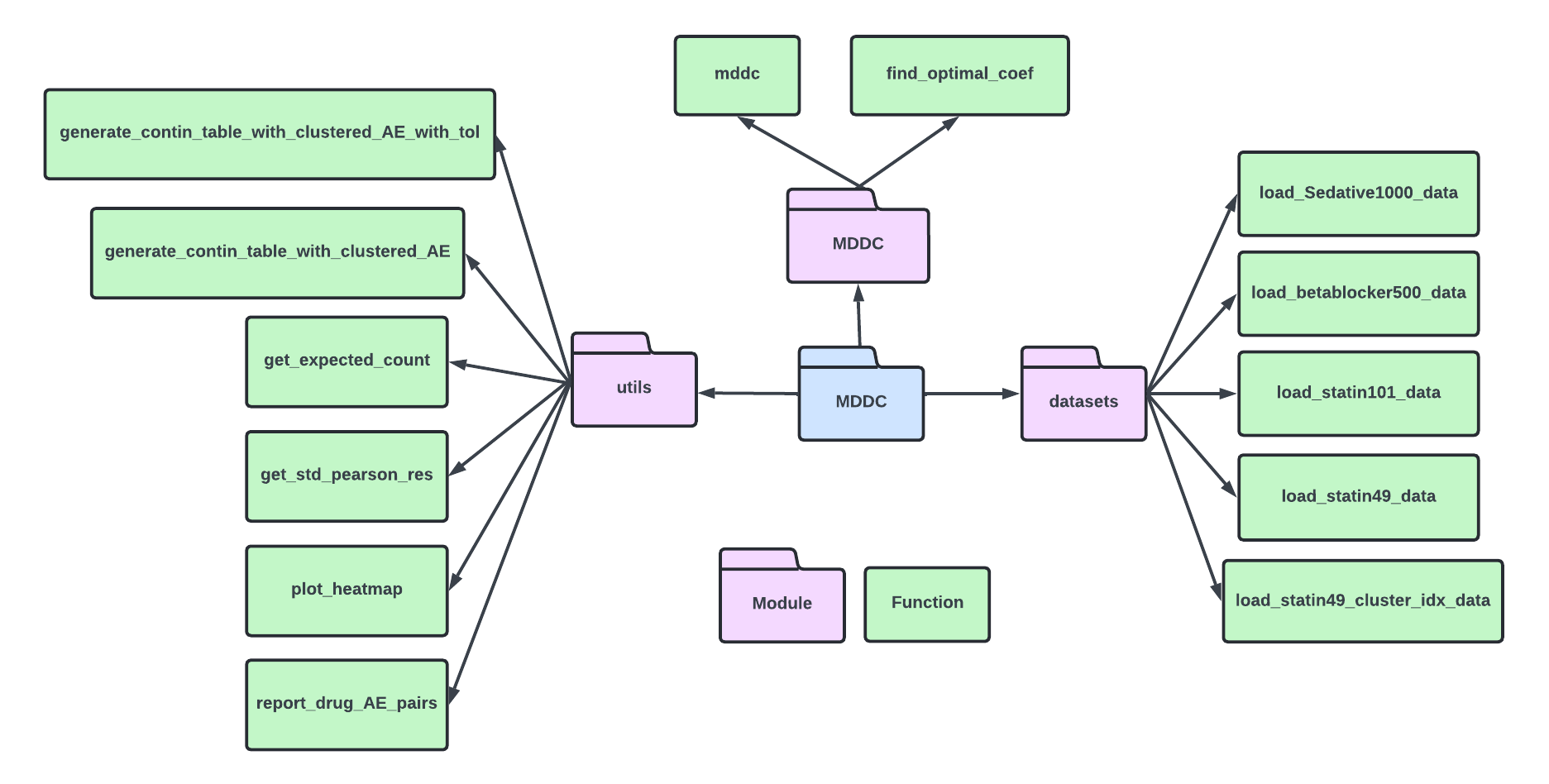}
    \caption{The modules and functions available in the \texttt{MDDC} in \texttt{Python}. The folder (blue) represents the \texttt{MDDC} package. The folders (lavender-colored) show the various modules within the package. The functions within the modules are depicted by the rectangles (green).}
    \label{fig:Py-MDDC}
\end{figure}

The structure of the \texttt{MDDC} in \texttt{Python} is depicted in Figure \ref{fig:Py-MDDC}. The \texttt{Python} package comprises of three modules, namely - \texttt{MDDC}, \texttt{datasets} and \texttt{utils}. The \texttt{MDDC} module contains the \texttt{mddc} function which implements both the Monte-Carlo and Boxplot methods for cutoff selection, and \texttt{find\_optimal\_coef}, for determining the optimal adaptive boxplot coefficient. The \texttt{utils} module contains the functions : \texttt{get\_expected\_count}, \texttt{get\_std\_pearson\_res}, \texttt{generate\_contin\_table\_with\_clustered\_AE}, \texttt{generate\_contin\_table\_with\_clustered\_AE\_with\_tol}, \texttt{report\_drug\_AE\_pairs} and \texttt{plot\_heatmap}, all of which offer the same functionality as their counterparts in the \texttt{R} package. Similarly, the \texttt{datasets} module includes functions to load five datasets (\texttt{load\_statin49\_data}, \texttt{load\_statin49\_cluster\_idx\_data}, \texttt{load\_statin101\_data}, \texttt{load\_betablocker500\_data}, and\\ \texttt{load\_sedative1000\_data}), comparable to those provided in the \texttt{R} package. The \texttt{Python} package can be found at \url{https://github.com/rmj3197/MDDC}, and the documentation at \url{https://mddc.readthedocs.io/en/latest/}. 

\subsection*{Salient Features}

Both the \texttt{R} and \texttt{Python} packages have been developed using best practices prevalent in the respective communities which ensures readability, maintainability, and compatibility with other tools and libraries. Additionally, to improve computational efficiency, both \texttt{R} and \texttt{Python} packages incorporate \texttt{C++}-based helper functions. The \texttt{R} and \texttt{Python} implementations use parallelization techniques to enhance the computational efficiency. Specifically, the \texttt{R} implementation utilizes the \texttt{doParallel} and \texttt{foreach} packages to achieve parallelization. Similarly, the \texttt{Python} implementation employs the \texttt{joblib} package to parallelized computationally intensive operations. We have also implemented continuous integration checks to ensure that code changes are automatically tested and integrated into the main codebase.

\section{Usage Examples}
\label{sec:example}

In this section, we present examples illustrating the usage of functions within the \texttt{MDDC} package. This package is designed to provide the MDDC algorithm with two data driven cutoff identification methods for evaluation of adverse event signals. Additionally it includes a data generation function for simulating pharmacovigilance datasets, and a few preprocessing and postprocessing functions. To install the package in \texttt{R}, use the following command:
\begin{verbatim}
> install.packages("MDDC")
\end{verbatim}
An analogue of the command in \texttt{Python} is:
\begin{verbatim}
pip install MDDC
\end{verbatim}
We start by loading the package in \texttt{R} using: 
\begin{verbatim}
> library(MDDC)
\end{verbatim}

We now describe the main functionalities provided by the \texttt{MDDC} package in both \texttt{R} and \texttt{Python}. As mentioned in Algorithm \ref{mddc_alg}, the packages provide two ways for determining the cutoff used in the algorithm, namely by using the upper limit of Tukey's Boxplot statistics (see Algorithm \ref{alg_boxplot_c}) and Monte Carlo Simulation (see Algorithm \ref{bootstrap_alg}). We demonstate the usage of the Boxplot method in \texttt{R} and Monte Carlo simulation method in \texttt{Python}. 

\subsection*{Dataset Description}

For illustrating the examples in this section, we will use a dataset on Beta Blockers which was created from the FAERS database, considering the period from the first quarter of 2021 (Q1 2021) to the fourth quarter of 2023 (Q4 2023). This dataset is a $501 \times 9$ contingency table. The first 500 rows represent 500 AEs with the highest reported marginals, while the final row aggregates the remaining AEs. 

Before starting the analysis, we begin by loading the data and examining the dataset:
\begin{verbatim}
> data("betablocker500")
> # showing the transposed head of the dataframe so
> # that it is not cutoff from the page
> t(head(betablocker500))

               Pain Fatigue Nausea Arthralgia Condition aggravated Dyspnoea
Acebutolol     3582       0     13         10                    1     1122
Atenolol        455     503    994        502                  474      305
Bisoprolol      977    1830   1313        892                  409     2368
Carvedilol      110     621    122        131                   90      549
Metoprolol      457    1191    816        617                  365     1702
Nadolol         307      34    349         17                    7       76
Propranolol     166     459   1023        188                  359      411
Timolol         348      79     51         42                  393      385
Other       1074063 1015237 757910     688797               679423   672028

                       
\end{verbatim}
Our goal is to identify (AE, drug) pairs with abnormally high report counts, specifically those cells with counts significantly exceeding their expected values.

\subsection*{AE Identification using the MDDC Boxplot Method}

The \texttt{mddc\_boxplot()} function in \texttt{R} implements the MDDC algorithm using the boxplot method for cutoff selection. This method can be used for its simplicity and computational benefits over the Monte Carlo method.


The \texttt{mddc\_boxplot()} function has the following arguments:
\begin{itemize}
    \item \texttt{contin\_table:} A data matrix of an $I \times J$ contingency table with rows representing adverse events and columns representing drugs. We recommend users first check the input contingency table using the function \texttt{check\_and\_fix\_ contin\_table()}.
    
    \item \texttt{col\_specific\_cutoff:} Logical. In the step 2 of the algorithm, whether to apply the boxplot method to the standardized Pearson residuals within each drug column (default is \texttt{TRUE}) or to the entire table (\texttt{FALSE}).
    
     \item \texttt{separate:} Logical. In the step 2 of the algorithm, whether to separate the standardized Pearson residuals for the zero cells and non zero cells and apply boxplot method separately or together. Default is \texttt{TRUE}.
    
    \item \texttt{if\_col\_cor:} Logical. In the step 3 of the algorithm, whether to use column (drug) correlation or row (adverse event) correlation. Default is \texttt{FALSE}, indicating the use of adverse event correlation. \texttt{TRUE} indicates the use of drug correlation.

    \item \texttt{cor\_lim:} A numeric value between 0 and 1. Specifies the correlation threshold to select ``connected'' adverse events in step 3. The default is 0.8.

    \item \texttt{coef:} A numeric value or a list of numeric values. If a single numeric value is provided, it will be applied uniformly across all columns of the contingency table. If a list is provided, its length must match the number of columns in the contingency table, and each value will be used as the coefficient for the corresponding column. Default is 1.5.

    \item \texttt{num\_cores:} Number of cores used to parallelize the MDDC Boxplot algorithm. Default is 2.
    
    \end{itemize}
We now perform the MDDC (boxplot) analysis with the \texttt{betablocker500} dataset:
    
\begin{verbatim}
> set.seed(42)
> boxplot_res <- mddc_boxplot(
+     contin_table = betablocker500, col_specific_cutoff = TRUE, separate = TRUE,
+     if_col_cor = FALSE, cor_lim = 0.8, num_cores = 8
+ )
\end{verbatim}
The above function outputs a list with three components:
\begin{itemize}
    \item \texttt{boxplot\_signal:} An $I\times J$ data matrix with entries 1 or 0, indicating the signals identified in the step 2. A value of 1 indicates signals, 0 indicates no signal.

    \item \texttt{corr\_signal\_pval:} An $I\times J$ data matrix of p-values for each cell in the contingency table from step 5, when the $r_{ij}$ values are mapped back to the standard normal distribution.

    \item \texttt{corr\_signal\_adj\_pval:} An $I\times J$ data matrix of the Benjamini-Hochberg adjusted p-values for each cell in step 5. Users can choose whether to use \texttt{corr\_signal\_pval} or \texttt{corr\_signal\_adj\_pval}, and can use their preferred p-value threshold for identifying signals. 
    \end{itemize}
    Below, we display the first few rows and columns for each component of \texttt{boxplot\_res}. We first check the component \texttt{boxplot\_signal}:
\begin{verbatim}
> # showing the transposed head of the dataframe so
> # that it is not cutoff from the page
> t(head(boxplot_res$boxplot_signal))
            Pain Fatigue Nausea Arthralgia Condition aggravated Dyspnoea
Acebutolol     1       0      0          0                    0        1
Atenolol       0       0      1          1                    1        0
Bisoprolol     0       0      0          0                    0        1
Carvedilol     0       1      0          0                    0        1
Metoprolol     0       0      0          0                    0        1
Nadolol        1       0      1          0                    0        0
Propranolol    0       0      1          0                    0        0
Timolol        1       0      0          0                    1        1
Other          0       0      0          0                    0        0
\end{verbatim}

From the above snippet, the identified signals in step 2 of MDDC (boxplot) include (Pain, Acebutolol), (Fatigue, Carvedilol), (Nausea, Atenolol), (Arthralgia, Atenolol), and (Condition aggravated, Atenolol). Additionally, Dyspnoea was associated with multiple drugs, including Acebutolol, Bisoprolol, Carvedilol, and Metoprolol. Now we look at the second component \texttt{corr\_signal\_pval} which shows the p-values of all the cells from step 5:

\begin{verbatim}
> # showing the transposed head of the dataframe so
> # that it is not cutoff from the page
> t(round(head(boxplot_res$corr_signal_pval), digits = 3))
            Pain Fatigue Nausea Arthralgia Condition aggravated Dyspnoea
Acebutolol    NA   0.703  0.665      0.647                   NA    0.385
Atenolol      NA   0.538  0.040      0.316                   NA    0.564
Bisoprolol    NA   0.535  0.593      0.604                   NA    0.008
Carvedilol    NA   0.261  0.934      0.877                   NA    0.052
Metoprolol    NA   0.547  0.605      0.627                   NA    0.001
Nadolol       NA   0.722  0.157      0.699                   NA    0.562
Propranolol   NA   0.736  0.119      0.817                   NA    0.587
Timolol       NA   0.795  0.741      0.752                   NA    0.075
Other         NA   0.411  0.422      0.407                   NA    0.955  
\end{verbatim}In this output, we observe that the first column, corresponding to the adverse event ``Pain'', does not have associated p-values. This is because, in step 2 of the algorithm, ``Pain'' is already identified as an signal for Acebutolol, Nadolol, and Timolol. Consequently, the standardized Pearson residual values for these three drugs were replaced with ``NA". For the remaining drugs, ``Pain" might have identified as an outlier with lower cells count, or more speciically with $e_{ij} < -c^{+}_{univ,j}$ which might have left only two or lesser residual values intact. With only two residual values remaining in the first row, it was not possible to find connected AEs for ``Pain''. Therefore, this adverse event was excluded from the subsequent steps of the analysis. Note that for computing Pearson correlation in step 3, at least three values are required in the matching positions. Applying a p-value threshold of 0.05, we identify the following pairs as signals by considering AE correlations. Nausea is identified as a signal for Atenolol, Dyspnoea is associated with both Bisoprolol and Metoprolol.  

The third component, \texttt{corr\_signal\_adj\_pval}, provides the Benjamini-Hochberg adjusted p-values. Users can choose whether to use \texttt{corr\_signal\_pval} or \texttt{corr\_signal\_adj\_pval} and can use p-value threshold of their preference e.g. 0.05.

\subsection*{AE Identification using the MDDC Monte Carlo Simulation Method}

Next, we introduce the Monte Carlo simulation method of MDDC with the corresponding function from the \texttt{Python} package. The \texttt{Python} function contains a single function \texttt{mddc} which can be used for both Boxplot and Monte Carlo simulation methods by specifying the \texttt{method} argument of the function. A brief overview of the most commonly used arguments for the function is provided below. Some advanced arguments are excluded for simplicity. If the reader is interested in greater details please see \url{https://mddc.readthedocs.io/en/latest/api_reference/generated/MDDC.MDDC.mddc.html}.

\begin{itemize} 
    \item \texttt{contin\_table}: An $I \times J$ contingency table where rows represent adverse events and columns represent drugs.

    \item \texttt{method}: Method for cutoff selection. Can be either \texttt{monte\_carlo} or \texttt{boxplot}.

    \item \texttt{rep}: Number of Monte Carlo replications used for estimating thresholds. Utilized in Step 2 of the MDDC algorithm. Only used if method is \texttt{monte\_carlo}.
    
    \item \texttt{quantile}: In step 2 of the algorithm, this specifies the quantile of the null distribution obtained via the Monte Carlo (MC) method to use as a threshold for identifying cells with high values of standardized Pearson residuals. The default is 0.95. Only used if method is \texttt{monte\_carlo}.
    
    \item \texttt{mc\_num}: The number of Monte Carlo replications to perform in step 2. The default is 10,000.
    
    \item \texttt{exclude\_same\_drug\_class}: In step 2, when applying Fisher's exact test to cells with a count less than six, a $2\times2$ contingency table is constructed. This argument specifies whether to exclude other drugs in the same class as the drug of interest. The default is \texttt{True}. Only used if method is \texttt{monte\_carlo}.
    
    \item \texttt{col\_specific\_cutoff}: Specifies whether to apply the specified method to the standardized Pearson residuals of the entire table or within each drug column in step 2. The default is \texttt{True}, indicating column-specific cutoff. \texttt{False} applies the method to the residuals of the entire table.
    
    \item \texttt{separate}: In step 2 of the algorithm, indicates whether to separate the standardized Pearson residuals for zero cells and non-zero cells, applying the specified method separately. The default is \texttt{TRUE}.
    
    \item \texttt{if\_col\_cor}: In step 3 of the algorithm, specifies whether to use column (drug) correlation or row (adverse event) correlation. The default is \texttt{False}, indicating the use of adverse event correlation. \texttt{True} indicates the use of drug correlation.
    
    \item \texttt{cor\_lim}: A numeric value between 0 and 1. Specifies the correlation threshold to use in step 3 for selecting ``connected'' adverse events. The default is 0.8.

    \item \texttt{n\_jobs}: This specifies the maximum number of concurrently running workers.

    \item \texttt{seed}: Seed for ensuring reproducibility while working with random number generators.
\end{itemize}

We now apply the MDDC (MC) algorithm to \texttt{statin49} using the following code:
\begin{verbatim}
>>> import MDDC
>>> betablocker500 = MDDC.datasets.load_betablocker500_data()
>>> mc_results = MDDC.MDDC.mddc(
...     contin_table=betablocker500,
...     method="monte_carlo",
...     rep=10000,
...     exclude_same_drug_class=True,
...     col_specific_cutoff=True,
...     separate=True,
...     if_col_corr=False,
...     corr_lim=0.8,
...     seed = 42
... )  
\end{verbatim}This function outputs a tuple with five elements:

\begin{itemize}
   \item \texttt{pval}: Returns the p-values for each cell in step 2. For cells with counts greater than five, the p-values are obtained via the Monte Carlo (MC) method. For cells with counts less than or equal to five, the p-values are obtained via Fisher's exact tests.
 \item \texttt{signal}: Indicates signals for cells with counts greater than five, identified in step 2 by the MC method. A value of 1 indicates a signal, while 0 indicates no signal.
 \item \texttt{fisher\_signal}: Indicates signals for cells with counts less than or equal to five, identified in step 2 by Fisher's exact tests. A value of 1 indicates a signal, while 0 indicates no signal.
 \item \texttt{corr\_signal\_pval}: Returns the p-values for each cell in the contingency table in step 5, where the \(r_{ij}\) values are mapped back to the standard normal distribution.
 \item \texttt{corr\_signal\_adj\_pval}: Returns the Benjamini-Hochberg adjusted p-values for each cell in step 5. Users can choose whether to use \texttt{corr\_signal\_pval} or \texttt{corr\_signal\_adj\_pval}, and select an appropriate p-value threshold (for example, 0.05). 
\end{itemize}

Both the \texttt{R}  and \texttt{Python} packages include a post-processing function \texttt{report\_drug\_AE\_pairs()} for displaying the identified (AE, drug) pairs as well as the observed count, expected count and the standardized Pearson residuals for the pairs. The function takes the arguments: 
\begin{itemize}
    \item \texttt{contin\_table:} A data matrix representing an $I \times J$ contingency table, with rows corresponding to adverse events and columns corresponding to drugs. 
\item \texttt{contin\_table\_signal:} A data matrix with the same dimensions and row and column names as \texttt{contin\_table}. Entries should be either 1 (indicating a signal) or 0 (indicating no signal). This matrix can be obtained by applying the \texttt{mddc\_boxplot()} or \texttt{mddc\_mc()} functions to \texttt{contin\_table}.
\end{itemize}

Now, we apply this function to the second element \texttt{signal} of the \texttt{mc\_results}. We display below the few rows of the output from the function. 
\begin{verbatim}
>>> MDDC.utils.report_drug_AE_pairs(betablocker500, mc_results.signal)[50:55]
        Drug                   AE  Observed Count  Expected Count  Std Pearson Resid
50  Atenolol       Liver disorder             209         44.1915            24.8128
51  Atenolol          Dehydration             124         42.7015            12.4514
52  Atenolol         Osteoporosis             337         42.2169            45.4060
53  Atenolol  Atrial fibrillation              84         40.7737             6.7749
54  Atenolol    Diabetes mellitus              73         38.4626             5.5732

\end{verbatim}
These (AE, drug) pairs are part of the signals identified by the MDDC (MC) method in step 2 for pairs with counts greater than five. Similarly, we can apply this function to the signals obtained from the correlation steps using the following code. Here we use a threshold of 0.05 for selecting the signals from step 5. We omit the output for brevity.

\begin{verbatim}
>>> MDDC.utils.report_drug_AE_pairs(betablocker500, (mc_results.pval <= 0.05))
\end{verbatim}

\begin{figure}[H]
    \centering
    \includegraphics[width=0.7\linewidth]{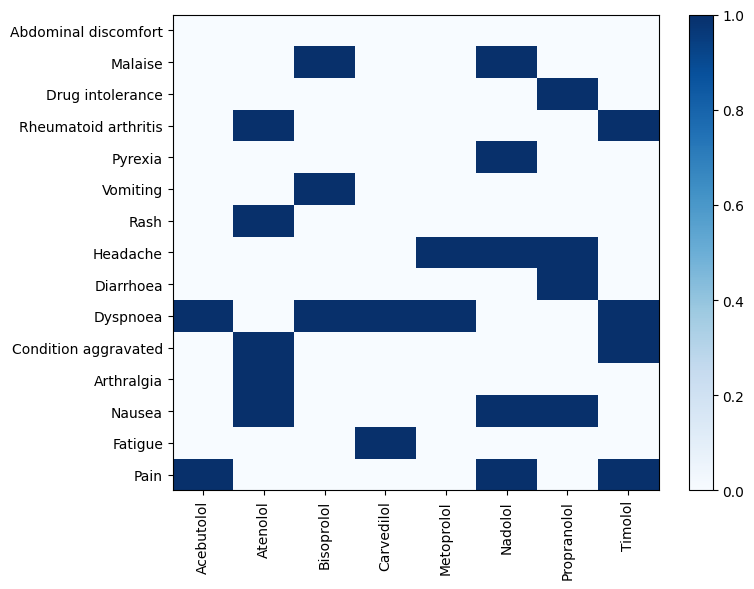}
    \caption{Heat map of a subset of the identified signals in step 2 of the algorithm using MDDC (MC).}
    \label{fig:heatmap}
\end{figure}

The \texttt{R} and \texttt{Python} packages also contain a function \texttt{plot\_heatmap} for plotting the heat map for the identified signals. A heatmap of the \texttt{signal} from \texttt{mc\_results} is shown in Figure \ref{fig:heatmap}.

\begin{verbatim}
>>> MDDC.utils.plot_heatmap(mc_results.signal.drop(columns=["Other"]).iloc[:15, :], 
... cmap= "Blues")
\end{verbatim}

\subsection*{Simulating datasets with grouped AEs}

The package offers a data generation function for simulating pharmacovigilance datasets, with the option to incorporate grouped AEs. This function can embed correlations between the standardized Pearson residuals for AEs and takes the following arguments:

\begin{itemize}
\item \texttt{row\_marginal}: Marginal sums for the rows of the contingency table.

\item \texttt{column\_marginal}:Marginal sums for the columns of the contingency table.

\item \texttt{signal\_mat}: A data matrix of the same dimensions as the contingency table with entries indicating the signal strength. Values must be greater than or equal to 1, where 1 indicates no signal, and values greater than 1 indicate a signal.

\item \texttt{contin\_table}: A data matrix representing an $I \times J$ contingency table with rows corresponding to adverse events and columns corresponding to drugs. The row and column marginals are used to generate the simulated data. 

\item \texttt{AE\_idx}: A data frame with two variables, \texttt{idx} and \texttt{AE}, where \texttt{idx} indicates the cluster index (either a name or a number), and \texttt{AE} lists the adverse event names. An example named \texttt{AE\_idx}, which provides the AE group index for the \texttt{statin49} dataset, is included in the package.

\item \texttt{n\_rep}: The number of simulated contingency tables to be generated.

\item \texttt{rho}: A numeric value representing the correlation of the AEs within each cluster. The default is 0.5.

\item \texttt{seed}: An optional integer to set the seed for reproducibility. 

\end{itemize}

Note that this function can either simulate tables based on the row and column marginals, or a contingency table. If a contingency table is available and the cluster indices and the value of $\rho$ are unknown, the function can estimate the correlation matrix based on the correlations between adverse events from the provided contingency table. The user also has the ability to explicitly specify a correlation matrix of choice as an input to the function. For demonstration purposes, we will generate three simulated datasets based on the marginals of \texttt{statin49} and the cluster indices specified in \texttt{statin49\_AE\_idx}.  

First, we need to create a data matrix with the same dimensions as \texttt{statin49} that indicates the signal strength for each (AE, drug) pair. In this example, we assign a signal (Rhabdomyolysis, Atorvastatin) with a strength of 4 to the simulated dataset:
\begin{verbatim}
> # create a matrix indicating signal strength
> sig_mat <- matrix(1,
+                   nrow = nrow(statin49),
+                   ncol = ncol(statin49))
> # assign (Rhabdomyolysis, Atorvastain) as a signal
> # with a signal strength 4
> sig_mat[1, 1] <- 4    
\end{verbatim}

The 49 AEs in \texttt{statin49} can be grouped into three clusters, as listed in the \texttt{AE\_idx}. Clutster 1 is for AEs associated with signs and symptoms of muscle injury, Cluster 2 is for AEs associated with laboratory tests for muscle injury, Cluster 3 for AEs associated with kidney injury and its laboratory diagnosis and treatment. Now we generate 3 simulated contingency tables based on the marginals of \texttt{statin49}, the pre-specified matrix of signal strength, and the AE group index, with a within group correlation $\rho=0.5$:

\begin{verbatim}
> row_marginal <- as.vector(rowSums(statin49))
> column_marginal <- as.vector(colSums(statin49))
> set.seed(42)
> sim_dat <- generate_contin_table_with_clustered_AE(
+     row_marginal = row_marginal,
+     column_marginal = column_marginal,
+     signal_mat = sig_mat,
+     AE_idx = statin49_AE_idx,
+     n_rep = 3,
+     rho = 0.5
+ )
\end{verbatim}

This function returns a list of simulated contingency tables, with the length of the list equal to the number of replications specified in the argument \texttt{n\_rep}. 

\section{Simulations}
\label{sec:performance}

\subsection*{Data Generation}


Algorithm \ref{data_generate_alg_new} presents a data generation algorithm with the ability to incorporate correlation among the different rows of an $I \times J$ table. The algorithm transforms standardized Pearson residuals sampled from a multivariate normal distribution with a general covariance matrix into counts. We modify this algorithm in order to generate tables that maintain the same total number of reports in each replication of a simulation study. 

To investigate the extent of the variation in the total number of reports in the simulated datasets using Algorithm \ref{data_generate_alg_new}, we performed a simulation study. We define a measure Relative Total Deviation (RTD) as 
$RTD = \dfrac{|n^{orig}_{\cdot \cdot} - n^{sim}_{\cdot \cdot}|}{n^{orig}_{\cdot \cdot}} \times 100$, where $n^{orig}_{\cdot \cdot}$ and $n^{sim}_{\cdot \cdot}$ denote the total number of reports in the original input and the simulated dataset. A simulation using various values of $n^{orig}_{\cdot \cdot}$, $I$, $J$, and $\rho$, and corresponding RTD values are shown in Table \ref{tab:data-gen}; 10,000 tables are generated for each setting and the corresponding statistics are shown. For the purposes of this simulation, no signal is embedded in any drug-AE pair i.e. $\lambda = 1$.

\begin{table}[]
\centering
\resizebox{0.65\columnwidth}{!}{%
\begin{tabular}{|c|c|c|c|lllll|}
\hline
\multirow{2}{*}{$n^{orig}_{\cdot \cdot}$} & \multirow{2}{*}{$I$} & \multirow{2}{*}{$J$} & \multirow{2}{*}{$\rho$} & \multicolumn{5}{c|}{\textbf{RTD}} \\ \cline{5-9} 
 &  &  &  & \multicolumn{1}{c|}{\textbf{Min.}} & \multicolumn{1}{c|}{\textbf{Median}} & \multicolumn{1}{c|}{\textbf{Mean}} & \multicolumn{1}{c|}{\textbf{Max.}} & \multicolumn{1}{c|}{\textbf{SD}} \\ \hline
\multirow{2}{*}{60} & \multirow{2}{*}{3} & \multirow{2}{*}{3} & 0 & \multicolumn{1}{l|}{0.000} & \multicolumn{1}{l|}{6.667} & \multicolumn{1}{l|}{6.959} & \multicolumn{1}{l|}{38.333} & 5.356 \\ \cline{4-9} 
 &  &  & 0.5 & \multicolumn{1}{l|}{0.000} & \multicolumn{1}{l|}{8.333} & \multicolumn{1}{l|}{9.765} & \multicolumn{1}{l|}{45.000} & 7.400 \\ \hline
\multirow{2}{*}{100} & \multirow{2}{*}{5} & \multirow{2}{*}{5} & 0 & \multicolumn{1}{l|}{0.000} & \multicolumn{1}{l|}{6.000} & \multicolumn{1}{l|}{6.565} & \multicolumn{1}{l|}{30.000} & 4.943 \\ \cline{4-9} 
 &  &  & 0.5 & \multicolumn{1}{l|}{0.000} & \multicolumn{1}{l|}{10.000} & \multicolumn{1}{l|}{11.204} & \multicolumn{1}{l|}{55.000} & 8.382 \\ \hline
1000 & 10 & 10 & 0 & \multicolumn{1}{l|}{0.000} & \multicolumn{1}{l|}{1.900} & \multicolumn{1}{l|}{2.276} & \multicolumn{1}{l|}{12.100} & 1.721 \\ \hline
\multirow{2}{*}{10000} & \multirow{2}{*}{100} & \multirow{2}{*}{100} & 0 & \multicolumn{1}{l|}{3.510} & \multicolumn{1}{l|}{7.090} & \multicolumn{1}{l|}{7.089} & \multicolumn{1}{l|}{10.640} & 0.916 \\ \cline{4-9} 
 &  &  & 0.5 & \multicolumn{1}{l|}{0.000} & \multicolumn{1}{l|}{7.125} & \multicolumn{1}{l|}{7.760} & \multicolumn{1}{l|}{32.650} & 5.144 \\ \hline
\multirow{3}{*}{100000} & 100 & 100 & \multirow{3}{*}{0} & \multicolumn{1}{l|}{0.000} & \multicolumn{1}{l|}{0.212} & \multicolumn{1}{l|}{0.251} & \multicolumn{1}{l|}{1.309} & 0.189 \\ \cline{2-3} \cline{5-9} 
 & 10 & 100 &  & \multicolumn{1}{l|}{0.000} & \multicolumn{1}{l|}{0.196} & \multicolumn{1}{l|}{0.235} & \multicolumn{1}{l|}{1.273} & 0.178 \\ \cline{2-3} \cline{5-9} 
 & 50 & 4 &  & \multicolumn{1}{l|}{0.000} & \multicolumn{1}{l|}{0.184} & \multicolumn{1}{l|}{0.216} & \multicolumn{1}{l|}{1.074} & 0.164 \\ \hline
1000000 & 100 & 100 & 0 & \multicolumn{1}{l|}{0.000} & \multicolumn{1}{l|}{0.067} & \multicolumn{1}{l|}{0.079} & \multicolumn{1}{l|}{0.392} & 0.060 \\ \hline
\multirow{2}{*}{57723334} & \multirow{2}{*}{102} & \multirow{2}{*}{5} & 0 & \multicolumn{1}{l|}{0.000} & \multicolumn{1}{l|}{0.001} & \multicolumn{1}{l|}{0.001} & \multicolumn{1}{l|}{0.004} & 0.001 \\ \cline{4-9} 
 &  &  & 1 & \multicolumn{1}{l|}{$3.464 \times 10^{-6}$} & \multicolumn{1}{l|}{0.028} & \multicolumn{1}{l|}{0.033} & \multicolumn{1}{l|}{0.163} & 0.025 \\ \hline
\end{tabular}%
}
\caption{Comparison of Original and Simulated Total Number of Reports Using RTD. The function \texttt{generate\_contin\_table\_with\_clustered\_AE} from \texttt{R} package was used for the simulation. Tables of dimension $I \times J$ were generated under the assumption of independence ($\rho = 0$) and with $\rho = 0.5, 1$ using Algorithm \ref{data_generate_alg_new}. In all cases, the covariance matrix is at least positive semi-definite.}
\label{tab:data-gen}
\end{table}

From table \ref{tab:data-gen}, we can observe the following aspects:
\begin{itemize}
    \item As $n^{orig}_{\cdot \cdot}$ increases, the RTD values generally become smaller. This suggests that if a user imposes strict conditions for lower RTD values in cases of smaller contingency tables, the algorithm may require a larger number of tries to generate the specified number of random contingency tables.
    \item For generating contingency tables with small dimensions and a limited number of reports, the RTD values exhibit a wide range of variation.

    \item In instances where a moderate number of reports (e.g. 10,000) are present in a large contingency table (e.g. $100 \times 100$), the RTD remains significantly large, though smaller than when the total number of reports is extremely small. This can be attributed to the higher number of rounding errors for a $100 \times 100$ contingency table and moderate number of total reports. 
    
    \item For contingency tables, with large total reports, the mean RTD values tend to decrease, and the variation (SD) becomes smaller. 
\end{itemize}

\noindent Taking these observations into account, we propose Algorithm \ref{data_generation_with_tol} for data generation which is a modification of the Algorithm \ref{data_generate_alg_new}. This algorithm is available in both the \texttt{R} and \texttt{Python} packages in the function\\ \texttt{generate\_contin\_table\_with\_clustered\_AE\_with\_tol}.  

\begin{breakablealgorithm}
\caption{A data generation algorithm considering correlated AEs incorporating tolerance for total report count.}
\label{data_generation_with_tol}
\begin{algorithmic}[1]
    \item Generate $n$ random contingency tables using Algorithm \ref{data_generate_alg_new}.
    \item Compute the RTD for each table as:
    \[
    RTD_i = \frac{|n^{\text{orig}}_{i \cdot \cdot} - n^{\text{sim}}_{i \cdot \cdot}|}{n^{\text{orig}}_{i \cdot \cdot}} \times 100
    \] where $i = 1, \ldots, n$.
    \item \textbf{If} $\max\limits_{n} [RTD_i] \leq {tol}$:
        \begin{itemize}
            \item All contingency tables are accepted.
        \end{itemize}
    \item \textbf{Else If} $\max\limits_{i} [RTD_i] > tol$:
    \begin{itemize}
            \item Identify all tables where $RTD_i > tol$.
            \item For each such contingency table:
            \begin{itemize}
                \item \textbf{While} $RTD_i > tol$:
                \begin{itemize}
                    \item Regenerate the contingency table using a different normal sample in Algorithm \ref{data_generate_alg_new}.
                \end{itemize}
            \end{itemize}
        \end{itemize}
\end{algorithmic}
\end{breakablealgorithm}


\subsection*{Performance}

We also conduct a brief simulation study to compare the time required for both Monte Carlo and Boxplot methods when applied to contingency tables of different sizes, using both \texttt{R} and \texttt{Python}. The simulation is performed on a macOS Sonoma system equipped with an Apple M1 processor running at 3.2 GHz with 16 GB of RAM. Additionally, the simulation is run using \texttt{Python} version 3.12.4 and \texttt{R} version 4.4.0. 

The simulated datasets are generated using the method described by Chakraborty et. al. (2022)\cite{Chakraborty2022} and implemented in the \texttt{R} package \texttt{pvLRT}\cite{pvlrt2023}. We generated 10 datasets for each configuration to assess the standard error associated with the computational time. 

\subsubsection*{Case I: No signals are embedded} 

The dimensions of the contingency tables used are outlined in Table \ref{tab:no-signal}. The row and column marginals from the datasets provided in the package are used for the simulation. Specifically, \texttt{statin101} for generating tables with dimensions $102 \times 5$, \texttt{betablocker500} with dimension a $501 \times 9$ table, and \texttt{sedative1000} with dimension $1001 \times 11$ table. The total number of reports ($n_{\cdot \cdot}$) corresponding to different dimensions are also reported. Moreover, groups of AEs are not considered in this study. The function \texttt{r\_contin\_table\_zip} from the \texttt{R} package \texttt{pvLRT}\cite{pvlrt2023} was used to generate random contingency tables with the specified row and column marginals. In this simulation, no signals were embedded in the dataset, meaning $\lambda_{ij} = 1$, and the zero inflation probabilities were set to zero as an input to the function.

\begin{table}[H]
\centering
\resizebox{0.7\columnwidth}{!}{%
\begin{tabular}{|c|c|cc|cc|}
\hline
 &  & \multicolumn{2}{c|}{\textbf{\begin{tabular}[c]{@{}c@{}}\texttt{R}\\ Time in secs [Mean (SD)]\end{tabular}}} & \multicolumn{2}{c|}{\textbf{\begin{tabular}[c]{@{}c@{}}\texttt{Python}\\ Time in secs [Mean (SD)]\end{tabular}}} \\ \cline{3-6} 
\multirow{-2}{*}{\textbf{Dimensions}} & \multirow{-2}{*}{\textbf{Total N. of Reports}} & \multicolumn{1}{c|}{\textbf{Boxplot}} & \textbf{Monte Carlo} & \multicolumn{1}{c|}{\textbf{Boxplot}} & \textbf{Monte Carlo} \\ \hline
\rowcolor[HTML]{D9EAD3} 
$102 \times 5$ & 57723334 & \multicolumn{1}{c|}{\cellcolor[HTML]{D9EAD3}0.181 (0.009)} & 1.659 (0.134) & \multicolumn{1}{c|}{\cellcolor[HTML]{D9EAD3}0.045 (0.013)} & 0.795 (0.015) \\ \hline
\rowcolor[HTML]{FFF2CC} 
$501 \times 9$ & 77367960 & \multicolumn{1}{c|}{\cellcolor[HTML]{FFF2CC}0.347 (0.040)} & 10.47 (0.705) & \multicolumn{1}{c|}{\cellcolor[HTML]{FFF2CC}0.265 (0.024)} & 5.759 (0.060) \\ \hline
\rowcolor[HTML]{F4CCCC} 
$1001 \times 11$ & 78030040 & \multicolumn{1}{c|}{\cellcolor[HTML]{F4CCCC}0.676 (0.023)} & 28.961 (0.616) & \multicolumn{1}{c|}{\cellcolor[HTML]{F4CCCC}0.547 (0.022)} & 13.782 (0.190) \\ \hline
\end{tabular}%
}
\caption{Execution times (in seconds) of \texttt{R} and \texttt{Python} packages for Boxplot and Monte Carlo methods across contingency tables of different dimensions when no signals are embedded.}
\label{tab:no-signal}
\end{table}

\subsubsection*{Case II: Signals are embedded in 30\% of the cells}

In this scenario, signals are randomly embedded in 30\% of the cells within each contingency table. Table \ref{tab:with-signal} presents the strengths of the embedded signals, along with the proportions of cells where the signals are embedded. Similar to Case I, the marginals from the datasets available in the package are used, along with the specified signal matrix, and zero inflation probabilities set to zero.

\begin{table}[H]
\centering
\resizebox{\columnwidth}{!}{%
\begin{tabular}{|c|c|c|cc|cc|}
\hline
 &  &  & \multicolumn{2}{c|}{\textbf{\begin{tabular}[c]{@{}c@{}}\texttt{R}\\ Time in secs [Mean (SD)]\end{tabular}}} & \multicolumn{2}{c|}{\textbf{\begin{tabular}[c]{@{}c@{}}\texttt{Python}\\ Time in secs [Mean (SD)]\end{tabular}}} \\ \cline{4-7} 
\multirow{-2}{*}{\textbf{Dimensions}} & \multirow{-2}{*}{\textbf{Total N. of Reports}} & \multirow{-2}{*}{\textbf{Signal Strengths and }} & \multicolumn{1}{c|}{\textbf{Boxplot}} & \textbf{Monte Carlo} & \multicolumn{1}{c|}{\textbf{Boxplot}} & \textbf{Monte Carlo} \\ \hline
\rowcolor[HTML]{D9EAD3} 
\cellcolor[HTML]{D9EAD3} & \cellcolor[HTML]{D9EAD3} & 4.5\% $\lambda$ = 2; 25.5\% $\lambda$ = 3 & \multicolumn{1}{c|}{\cellcolor[HTML]{D9EAD3}0.381 (0.046)} & 1.621 (0.156) & \multicolumn{1}{c|}{\cellcolor[HTML]{D9EAD3}0.060 (0.008)} & 0.774 (0.017) \\ \cline{3-7} 
\rowcolor[HTML]{D9EAD3} 
\cellcolor[HTML]{D9EAD3} & \cellcolor[HTML]{D9EAD3} & 9\% $\lambda$ = 2; 21\% $\lambda$ = 3 & \multicolumn{1}{c|}{\cellcolor[HTML]{D9EAD3}0.843 (1.209)} & 1.725 (0.124) & \multicolumn{1}{c|}{\cellcolor[HTML]{D9EAD3}0.068 (0.017)} & 0.789 (0.019) \\ \cline{3-7} 
\rowcolor[HTML]{D9EAD3} 
\cellcolor[HTML]{D9EAD3} & \cellcolor[HTML]{D9EAD3} & 4.5\% $\lambda$ = 2; 25.5\% $\lambda$ = 4 & \multicolumn{1}{c|}{\cellcolor[HTML]{D9EAD3}0.467 (0.115)} & 1.696 (0.113) & \multicolumn{1}{c|}{\cellcolor[HTML]{D9EAD3}0.062 (0.009)} & 0.784 (0.018) \\ \cline{3-7} 
\rowcolor[HTML]{D9EAD3} 
\cellcolor[HTML]{D9EAD3} & \cellcolor[HTML]{D9EAD3} & 9\% $\lambda$ = 2; 21\% $\lambda$ = 4 & \multicolumn{1}{c|}{\cellcolor[HTML]{D9EAD3}0.450 (0.112)} & 1.684 (0.081) & \multicolumn{1}{c|}{\cellcolor[HTML]{D9EAD3}0.064 (0.010)} & 0.773 (0.021) \\ \cline{3-7} 
\rowcolor[HTML]{D9EAD3} 
\cellcolor[HTML]{D9EAD3} & \cellcolor[HTML]{D9EAD3} & 4.5\% $\lambda$ = 3; 25.5\% $\lambda$ = 4 & \multicolumn{1}{c|}{\cellcolor[HTML]{D9EAD3}0.655 (0.767)} & 1.660 (0.098) & \multicolumn{1}{c|}{\cellcolor[HTML]{D9EAD3}0.064 (0.009)} & 0.771 (0.015) \\ \cline{3-7} 
\rowcolor[HTML]{D9EAD3} 
\cellcolor[HTML]{D9EAD3} & \cellcolor[HTML]{D9EAD3} & 9\% $\lambda$ = 3; 21\% $\lambda$ = 4 & \multicolumn{1}{c|}{\cellcolor[HTML]{D9EAD3}0.434 (0.064)} & 1.709 (0.116) & \multicolumn{1}{c|}{\cellcolor[HTML]{D9EAD3}0.058 (0.010)} & 0.776 (0.019) \\ \cline{3-7} 
\rowcolor[HTML]{D9EAD3} 
\multirow{-7}{*}{\cellcolor[HTML]{D9EAD3}$102 \times 5$} & \multirow{-7}{*}{\cellcolor[HTML]{D9EAD3}57723334} & 10\% $\lambda$ = 2; 10\% $\lambda$ = 3; 10\% $\lambda$ = 4 & \multicolumn{1}{c|}{\cellcolor[HTML]{D9EAD3}0.377 (0.045)} & 1.656 (0.124) & \multicolumn{1}{c|}{\cellcolor[HTML]{D9EAD3}0.064 (0.016)} & 0.785 (0.020) \\ \hline
\rowcolor[HTML]{FFF2CC} 
\cellcolor[HTML]{FFF2CC} & \cellcolor[HTML]{FFF2CC} & 4.5\% $\lambda$ = 2; 25.5\% $\lambda$ = 3 & \multicolumn{1}{c|}{\cellcolor[HTML]{FFF2CC}6.473 (1.058)} & 13.062 (0.531) & \multicolumn{1}{c|}{\cellcolor[HTML]{FFF2CC}1.174 (0.139)} & 6.147 (0.127) \\ \cline{3-7} 
\rowcolor[HTML]{FFF2CC} 
\cellcolor[HTML]{FFF2CC} & \cellcolor[HTML]{FFF2CC} & 9\% $\lambda$ = 2; 21\% $\lambda$ = 3 & \multicolumn{1}{c|}{\cellcolor[HTML]{FFF2CC}6.274 (0.655)} & 13.172 (0.420) & \multicolumn{1}{c|}{\cellcolor[HTML]{FFF2CC}1.121 (0.119)} & 6.143 (0.094) \\ \cline{3-7} 
\rowcolor[HTML]{FFF2CC} 
\cellcolor[HTML]{FFF2CC} & \cellcolor[HTML]{FFF2CC} & 4.5\% $\lambda$ = 2; 25.5\% $\lambda$ = 4 & \multicolumn{1}{c|}{\cellcolor[HTML]{FFF2CC}8.750 (0.931)} & 14.081 (0.529) & \multicolumn{1}{c|}{\cellcolor[HTML]{FFF2CC}1.423 (0.138)} & 6.194 (0.099) \\ \cline{3-7} 
\rowcolor[HTML]{FFF2CC} 
\cellcolor[HTML]{FFF2CC} & \cellcolor[HTML]{FFF2CC} & 9\% $\lambda$ = 2; 21\% $\lambda$ = 4 & \multicolumn{1}{c|}{\cellcolor[HTML]{FFF2CC}7.426 (1.101)} & 13.781 (0.606) & \multicolumn{1}{c|}{\cellcolor[HTML]{FFF2CC}1.157 (0.143)} & 6.175 (0.140) \\ \cline{3-7} 
\rowcolor[HTML]{FFF2CC} 
\cellcolor[HTML]{FFF2CC} & \cellcolor[HTML]{FFF2CC} & 4.5\% $\lambda$ = 3; 25.5\% $\lambda$ = 4 & \multicolumn{1}{c|}{\cellcolor[HTML]{FFF2CC}10.745 (1.471)} & 13.804 (0.614) & \multicolumn{1}{c|}{\cellcolor[HTML]{FFF2CC}1.521 (0.159)} & 6.189 (0.066) \\ \cline{3-7} 
\rowcolor[HTML]{FFF2CC} 
\cellcolor[HTML]{FFF2CC} & \cellcolor[HTML]{FFF2CC} & 9\% $\lambda$ = 3; 21\% $\lambda$ = 4 & \multicolumn{1}{c|}{\cellcolor[HTML]{FFF2CC}9.865 (1.076)} & 13.728 (0.705) & \multicolumn{1}{c|}{\cellcolor[HTML]{FFF2CC}1.407 (0.153)} & 6.188 (0.107) \\ \cline{3-7} 
\rowcolor[HTML]{FFF2CC} 
\multirow{-7}{*}{\cellcolor[HTML]{FFF2CC}$501 \times 9$} & \multirow{-7}{*}{\cellcolor[HTML]{FFF2CC}77367960} & 10\% $\lambda$ = 2; 10\% $\lambda$ = 3; 10\% $\lambda$ = 4 & \multicolumn{1}{c|}{\cellcolor[HTML]{FFF2CC}5.713 (0.911)} & 13.456 (0.640) & \multicolumn{1}{c|}{\cellcolor[HTML]{FFF2CC}1.015 (0.119)} & 6.150 (0.120) \\ \hline
\rowcolor[HTML]{F4CCCC} 
\cellcolor[HTML]{F4CCCC} & \cellcolor[HTML]{F4CCCC} & 4.5\% $\lambda$ = 2; 25.5\% $\lambda$ = 3 & \multicolumn{1}{c|}{\cellcolor[HTML]{F4CCCC}19.550 (2.802)} & 34.805 (0.789) & \multicolumn{1}{c|}{\cellcolor[HTML]{F4CCCC}3.071 (0.278)} & 14.620 (0.191) \\ \cline{3-7} 
\rowcolor[HTML]{F4CCCC} 
\cellcolor[HTML]{F4CCCC} & \cellcolor[HTML]{F4CCCC} & 9\% $\lambda$ = 2; 21\% $\lambda$ = 3 & \multicolumn{1}{c|}{\cellcolor[HTML]{F4CCCC}15.493 (1.378)} & 34.225 (0.875) & \multicolumn{1}{c|}{\cellcolor[HTML]{F4CCCC}2.664 (0.164)} & 14.677 (0.334) \\ \cline{3-7} 
\rowcolor[HTML]{F4CCCC} 
\cellcolor[HTML]{F4CCCC} & \cellcolor[HTML]{F4CCCC} & 4.5\% $\lambda$ = 2; 25.5\% $\lambda$ = 4 & \multicolumn{1}{c|}{\cellcolor[HTML]{F4CCCC}26.098 (2.898)} & 36.474 (0.826) & \multicolumn{1}{c|}{\cellcolor[HTML]{F4CCCC}4.023 (0.409)} & 14.865 (0.403) \\ \cline{3-7} 
\rowcolor[HTML]{F4CCCC} 
\cellcolor[HTML]{F4CCCC} & \cellcolor[HTML]{F4CCCC} & 9\% $\lambda$ = 2; 21\% $\lambda$ = 4 & \multicolumn{1}{c|}{\cellcolor[HTML]{F4CCCC}22.089 (3.327)} & 37.302 (1.843) & \multicolumn{1}{c|}{\cellcolor[HTML]{F4CCCC}3.188 (0.422)} & 14.724 (0.379) \\ \cline{3-7} 
\rowcolor[HTML]{F4CCCC} 
\cellcolor[HTML]{F4CCCC} & \cellcolor[HTML]{F4CCCC} & 4.5\% $\lambda$ = 3; 25.5\% $\lambda$ = 4 & \multicolumn{1}{c|}{\cellcolor[HTML]{F4CCCC}28.427 (3.216)} & 36.546 (0.832) & \multicolumn{1}{c|}{\cellcolor[HTML]{F4CCCC}4.347 (0.426)} & 14.726 (0.237) \\ \cline{3-7} 
\rowcolor[HTML]{F4CCCC} 
\cellcolor[HTML]{F4CCCC} & \cellcolor[HTML]{F4CCCC} & 9\% $\lambda$ = 3; 21\% $\lambda$ = 4 & \multicolumn{1}{c|}{\cellcolor[HTML]{F4CCCC}26.694 (2.856)} & 36.497 (0.700) & \multicolumn{1}{c|}{\cellcolor[HTML]{F4CCCC}4.196 (0.450)} & 14.763 (0.219) \\ \cline{3-7} 
\rowcolor[HTML]{F4CCCC} 
\multirow{-7}{*}{\cellcolor[HTML]{F4CCCC}$1001 \times 11$} & \multirow{-7}{*}{\cellcolor[HTML]{F4CCCC}78030040} & 10\% $\lambda$ = 2; 10\% $\lambda$ = 3; 10\% $\lambda$ = 4 & \multicolumn{1}{c|}{\cellcolor[HTML]{F4CCCC}15.121 (2.357)} & 34.911 (0.736) & \multicolumn{1}{c|}{\cellcolor[HTML]{F4CCCC}2.716 (0.332)} & 14.575 (0.347) \\ \hline
\end{tabular}%
}
\caption{Execution times (in seconds) of \texttt{R} and \texttt{Python} packages for Boxplot and Monte Carlo methods across contingency tables of different dimensions when signals are embedded in 30\% of the cells.}
\label{tab:with-signal}
\end{table}

We can observe from Table \ref{tab:no-signal} and Table \ref{tab:with-signal}, that the presence of a signal in the contingency table impacts the execution times of the Boxplot method considerably, in case of both \texttt{R} and \texttt{Python} implementations. 
In the case of the Monte Carlo method, the difference in time between the two cases is not significant. Furthermore, the relative differences in execution times between the \texttt{R} and \texttt{Python} implementations suggest that both approaches are efficient and perform comparably well. A particularly notable observation is the significantly lower computational time required by the Boxplot method in both the implementations. It is important to emphasize that while the Monte-Carlo method is more computationally intensive, it is also more robust in terms of better control of false discovery rate and higher specificity, whereas the Boxplot method is intended to serve primarily as an exploratory approach.

Additionally, it is important to note that while the execution time for the Monte Carlo method may increase with larger contingency tables, in practical scenarios, contingency tables of arbitrarily large dimensions are rarely used. Typically, AEs are assessed within specific classes of drugs or in relation to a particular class of serious AEs, which leads to construction of contingency tables which are well curated and are of lower dimension.

\section{Discussion and Future Developments}

We presented software, developed in \texttt{R} and \texttt{Python}, that encodes a novel pattern discovery algorithm with statistical guarantees for AE identification. The algorithm, named MDDC, identifies AEs as outliers in the cells of an $I \times J$ contingency table.

The MDDC algorithm has its roots in the field of robustness, as it is based on the cell wise contamination model introduced by Alqallaf et. al. (2009) \cite{alqallaf2009propagation}. This model considers that each cell in a data matrix can be contaminated individually and independently of the remaining cells in the same row. In our pharmacovigilance framework, this means that a potential signal may be an AE for one drug and not for another. The case where a signal is an AE for all drugs in a class is covered by step 2 of the algorithm. An important aspect of the algorithm is the identification of AEs by considering the dependence among two rows of the table. We measure the dependence between the two rows of an $I \times J$ table by the Pearson correlation that is computed between the two rows of Pearson residuals. The question of whether Pearson correlation provides the best measure of dependence between two rows/columns of an $I \times J$ table is debatable and more research is needed to answer this question.

A second important aspect of the algorithm is the identification of appropriate cutoffs that are based on statistical performance guarantees. We proposed algorithms that are based on controlling the FDR. We also discussed a data generation method that is based on Pearson residuals and incorporates correlation among these residuals, and hence among the rows of the table.

Several packages encode methods based on tests for AE identification, but the study of pattern discovery in pharmacovigilance is insufficiently developed. An exception is TreeScan. This method depends on medical ontologies such as Medical Dictionary for Regulatory Activities (MedDRA). One prominent characteristic, common in these ontologies is their hierarchical structure. The TreeScan requires the user to supply the initial tree of relationships associated with the potential AE. This is problematic, especially in the case of novel AEs for which these relationships are unknown. The MDDC algorithm is independent of medical ontologies and does not require one to supply any additional information beyond what is observed. As such, it can be readily applied to the observed data. Finally, our implementation shows that the execution time is generally fast. 

\label{sec:discussion}

\section*{Data Availability Statement}

The data that are used in this manuscript are available in the MDDC \texttt{R} and \texttt{Python} packages, which can be found at \url{https://github.com/niuniular/MDDC} and \url{https://github.com/rmj3197/MDDC} respectively.

\section*{Acknowledgements}

This work was partially funded by FDA (contract number 75F40120C00159) and by Kaleida Health Foundation awards to the last author. 
The authors would like to thank Drs Robert Ball and Oanh Dang, FDA for stimulating discussions.

\section*{Author contributions statement}

A.L. developed the \texttt{R} package and contributed to the writing of the manuscript. R.M. developed the \texttt{Python} package and contributed to the writing of the manuscript. M.M. obtained funding, contributed to the writing of the manuscript, and supervised the work presented here. All authors reviewed the manuscript. 

\section*{Competing Interests}


The authors have no competing interests.

\end{document}